\documentstyle[12pt]{article}
\begin{document}
\begin{titlepage}

 \vspace*{.15in}
\begin{center}
{\bf\large
The Evolution of the Pion Distribution Amplitude \\
in Next-to-Leading Order}
\vspace*{0.35in}

D. M\"uller  \\
Stanford Linear Accelerator Center, \\
       Stanford University, Stanford, California  94309
\end{center}

\vspace*{0.25in}

\begin{abstract}
The evolution of the pion distribution amplitude in next-to-leading order is
studied for a fixed and a running coupling constant. In both cases, the
evolution provides a logarithmic modification in the endpoint region.
Assuming a simple parameterization of the distribution amplitude at a scale of
$Q_0\sim 0.5\ \rm GeV$,  it is shown numerically that these effects are large
enough at $Q\sim 2\ \rm  GeV$ that they have to be taken into account in the
next-to-leading-order analysis for exclusive processes. Alternatively, by
introducing a new distribution amplitude that evolves more smoothly,  this
logarithmic modification can be included in the hard-scattering part of the
considered process.
\end{abstract}
\vspace*{.25in}



\section{Introduction}

The perturbative approach for hard exclusive quantum chromodynamic (QCD)
processes was developed for more than one decade
\cite{BroLep79a,CeZy77,EfrRad80,DucMue80} (see Ref.~\cite{rewvExscl} for
reviews). In this approach, the scattering amplitude at large momentum transfer
$Q^2$  factorizes as a convolution of process-independent distribution
amplitudes, with a process-dependent perturbatively computable hard-scattering
amplitude. By using the leading-order perturbative QCD (pQCD) analysis,
 which was performed
for a large number of exclusive processes including mesons and baryons,
the qualitatively behavior for large $Q^2$ could be well understood
\cite{BroLep80,Bro93}.
However, using the asymptotic distribution amplitudes, which follow directly
from the solution of the evolution equation, results in predicted
normalizations for the elastic form factors at experimental accessible momentum
transfer that are too small;  in the case of the magnetic nucleon form factor,
this  provides the opposite sign.

{}From deep inelastic scattering, where the application of pQCD is
generally accepted, it is known that the used parton distribution
functions for accessible $Q^2$ are far from their asymptotic form
where all higher moments $m_n$, i.e., $n>0$, vanish.
It is therefore expected that for the exclusive processes
 at accessible momentum transfer, the distribution amplitudes
are nonasymptotical. Choosing distribution amplitudes that
are enhanced in the endpoint region (and asymmetric for nucleons)
 provides the observed normalization and sign for the elastic
form factors.

\end{titlepage}

 Reference\ \cite{IsSm}  argues that choosing such enhanced amplitudes
provides inconsistencies that affect the importance of higher twist
contributions,  as well as of perturbative nonleading-order terms, and so the
pQCD approach to elastic form factors probably is not self-consistent.
(A second point widely discussed  in the literature is the nonperturbative
contribution from the hadronic wave function \cite{IsSm,Radcri}.)
Phenomenological methods,  such as  (1)  introduce a gluon mass,
(2) freeze the running coupling constant for small virtuality \cite{phst1}, or
(3) suppress  the endpoint region by suitable distribution amplitudes or by a
cutoff \cite{phst2},  are used  to improve the stability of the pQCD approach.
Recent incorporation of Sudakov suppression has shown that the pQCD
 approach for the pion form factor is self-consistent for a momentum transfer
of $Q\sim 20\Lambda_{QCD}$ \cite{LiSte92} (see also Ref.\ \cite{JakKro94}).

The validity of the pQCD approach for
exclusive processes can also be studied by direct calculations of higher twist
and perturbative nonleading contributions.
It appears that higher twist analyses have not  been achieved
quantitatively.
The stability of the perturbation theory has been investigated
neglecting the evolution of the distribution amplitude
by next-to-leading-order  calculations for the pion transition form factor
\hbox{Ref. \cite{AguCha81,Bra83},}  the pion form factor
\cite{FieGupOttCha81,DitRad81}, and  the two-photon processes
\hbox{$\gamma\gamma\to M^+ M^- (M=\pi,K)$ \cite{Niz87}.}
Discrepancies in the one-loop approximation of the hard-scattering amplitude
for the pion form factor were clarified in Ref.\ \cite{KhaRad85,BraTse87}.
The next-to-leading-order correction to the pion form factor and to the
processes $\gamma\gamma\to M^+ M^-$  are rather large at accessible momentum
transfer.

Including the evolution of the
distribution amplitude in these analyses requires the solution of the
differential-integral evolution equation, which can be done by using the
moment method.
The corresponding two-loop approximation of the integral kernel was computed
by different authors and the obtained results agree with each other
\cite{DitRad/Sar84/Kat/MikRad85}.
It has been confirmed that the computed evolution kernel is
consistent with the Gribov-Lipatov-Altarelli-Parisi kernel \cite{GeDiHoMuRo}
and with conformal symmetry breaking in massless gauge field theories
\cite{DM94}.
Because of the complicated structure of the evolution kernel, only the first
few moments of the evolution kernel had been computed numerically
\cite{MiRa86}. Based on this incomplete computation,  it was believed that the
next-to-leading-order correction to the evolution of the distribution amplitude
and the contribution of this correction to the pion
form factor are rather small~\cite{MiRa86,KaMiRa86}.

Recently, using conformal constraints,  the complete formal solution of the
evolution equation in next-to-leading order could be obtained
 without knowing the evolution
kernel by a one-loop calculation \cite{DM94}.
This paper studies this solution in detail
and shows that the evolution of the
distribution amplitude must be included in
the next-to-leading order analysis.
 Section 2 reviews to leading order the evolution equation
 of the distribution amplitude and the
solution in terms of the conformal spin expansion.
The evolution of the distribution amplitude in next-to-leading order
for fixed $\alpha_s$ is  studied in Section 3.
This includes a detailed investigation of the large $n$ behavior for the
next-to-leading order corrections to the eigenfunctions
$\varphi^{ef}_n(x,\alpha_s)$
and eigenvalues $\gamma_n(\alpha_s)$ of the evolution kernel.
Numerical results for the evolution of the asymptotic,
the Chernyak-Zhitnitsky two-hump, and another convex
distribution amplitude are presented.
 Section 4 analyzes the solution of the evolution equation in
next-to-leading order with  running coupling,
showing  by numerical computation  that
the next-to-leading-order corrections are also large in this case.
 Section 5 discusses the obtained result,  comparing it with a previous
result \cite{MiRa86}, and presents the conclusions.

\section{The Distribution Amplitude and their Evolution}

The distribution amplitude $\varphi(x,Q^2)$ is the
probability amplitude for finding a valence quark [antiquark] with light
cone momentum fraction $x$ [$1-x$] in the pion probed at large momentum
square $Q^2$ \cite{BroLep79a}.
This amplitude can be defined as expectation value of renormalized nonlocal
light cone operators \cite{BroLep79a,GeDiHoMuRo}
\begin{eqnarray}
\label{def-DA}
\varphi(x,Q^2)\ =\ f_\pi^{-1}    \int {d\kappa\over \pi}\
            \exp\left[i\kappa (\tilde nP)\,(2x-1)\right]\
 \left.\vphantom{)_)}
       \left\langle0|O(\kappa;\tilde n)|P\right\rangle
\,\right\vert_{\textstyle \,\mu^2=Q^2} \ \ ,
\end{eqnarray}
where for simplicity, the renormalization point $\mu^2$ is set equal to the
large momentum transfer $Q^2$  (this choice is not optimal with respect to the
factorization scale setting for the considered processes;  however, it is
sufficient for the following discussions).
The light ray vector $\tilde n$ is chosen as
$\tilde n=(\tilde n_+=0,\tilde n_-=2,\vec{0}_\bot)$ so that $\tilde n P=P_+$,
 $\left. |P\right\rangle$ denotes the pion state with momentum $P$, and
\begin{eqnarray}
\label{def-LCO}
O(\kappa;\tilde n)\ =\   :\!\bar \psi_d(-\kappa\tilde n)\,  \gamma_5 (\tilde  n
\gamma)\,  U(-\kappa\tilde n,\kappa\tilde n)\,  \psi_u(\kappa\tilde n)\!:
\end{eqnarray}
is the light-cone operator with the flavor content of the considered pion.
The path ordered phase factor $U(-\kappa\tilde n,\kappa\tilde n)$
ensures the gauge invariance of this operator.
The pion decay constant  $f_\pi=133$ MeV introduced in (1)
guarantees the normalization \cite{BroLep79a}
\begin{eqnarray}
\label{sumrule}
\int_0^1 dx\  \varphi(x,Q^2)\ \ =\
\left.{f_\pi^{-1}
\left\langle 0|:\!\bar\psi_d(0)\,\gamma_5(\tilde n\gamma)\,\psi_u(0)\!:|P
\right\rangle}
\vphantom{_0^1}\right\slash
{\tilde n P} \  =\ 1\ .
\end{eqnarray}
Analogous to a quantum mechanical ground state,  it is to be expected that
$\varphi(x,Q^2)$ can be chosen positive.
Notice that because of charge conjugation invariance,  the symmetry
relation $\varphi(x,Q^2)=\varphi(1-x,Q^2)$ holds true.

The evolution equation for $\varphi(x,Q^2)$  derived in
Ref.~\cite{BroLep79a,BroLep80} can also be obtained in a straightforward
manner from the renormalization
group equation of the nonlocal operator $O(\kappa;\tilde n)$ \cite{GeDiHoMuRo}
\begin{eqnarray}
\label{eveqda}
Q^2{d\over dQ^2}\ \varphi(x,Q^2)\ =\ \int_0^1 dy V
\left( x,y; \alpha_s(Q^2)\right)\ \varphi(y,Q^2)\ ,
\end{eqnarray}
where $\alpha_s=g^2/(4\pi)$ is the QCD fine structure constant.
The evolution kernel
 $V(x,y;\alpha_s)=
(\alpha_s / 2\pi)\ V^{(0)}(x,y)+(\alpha_s / 2\pi)^2\ V^{(1)}(x,y)+\cdots\ $
has been computed perturbatively in  one-- and  two--loop  approximation
by using  the dimensional regularization in the modified minimal subtracted
$\overline{\mbox{MS}}$-scheme \cite{DitRad/Sar84/Kat/MikRad85}.
\goodbreak

The evolution equation (\ref{eveqda}) can be solved by conformal spin
expansion
\begin{eqnarray}
\label{dadeopco}
\varphi(x,Q^2)\ =\ \sum_{n=0}^{\infty}\!^{^{\scriptstyle \prime}}\
{{(1-x)x} \over N_n}\   C_n^{3/ 2}(2x-1)\
\left\langle 0| O_{n}(\mu^2)|P\right\rangle^{\rm red}_{\left.\vphantom{)_)}
\right\vert_{\textstyle \, \mu^2=Q^2}}\ ,
\end{eqnarray}
$$
 N_n\ =\ {(n+1)\,(n+2)\over 4(2n+3)}\ ,\phantom{MMMMMMMM}
$$
where the sum runs only over even $n$ [to ensure the above mentioned
symmetry of $\varphi(x,Q^2)$].
Here,  $\left\langle 0| O_{n}(Q^2)|P\right\rangle^{\rm red}
=\int_0^1 dx C_n^{3/2}(2x-1)\varphi(x,Q^2)$ are
reduced expectation values of local operators that in leading order do not
mix under renormalization \cite{EfrRad80,BroFriLepSac80}.
In the free field theory, these operators labeled by the conformal spin
 form an infinite irreducible representation of the so-called collinear
conformal algebra, which is a
subalgebra $O(2,1)$ of the full conformal algebra $O(4,2)$
\cite{Mak/Ohr/DobTod}. The Gegenbauer polynomials $C_n^{3/2}$ of order $3/2$
form an orthogonal and complete basis in the space of quadrate integrable
functions with the weight $(1-x)x$.
Thus, expansion (\ref{dadeopco}) converges if $\varphi(x,Q^2)$ vanishes at
the endpoints of the interval $[0,1]$; see, for instance,
Ref.~\cite{BatErd53appr}.
This condition is automatically satisfied \cite{BroLep80}.
 \goodbreak

The $Q$-dependence of $\left\langle 0|O_{n}(Q^2)|P\right\rangle^{\rm red}$
can be determined from the evolution equation
\begin{eqnarray}
\label{opcorgeq}
Q^2\ {d\over d Q^2}\
\left\langle 0|O_{n}(Q^2)|P\right\rangle\  =\
 {1\over 2}\
\sum_{k=0}^n\!^{^{\scriptstyle \prime}}\
 \gamma_{nk}\big(\alpha_s(Q^2)\big)\
\left\langle 0|O_{k}(Q^2)|P\right\rangle\ ,
\end{eqnarray}
where the anomalous dimension matrix
$\gamma_{nk}=(\alpha_s/2\pi)\ \gamma^{(0)}_n\delta_{nk}+
              (\alpha_s/2\pi)^2\gamma^{(1)}_{nk}+\cdots\ $
is diagonal in one--loop order.
In general, Poincar\'e-invariance of the theory assures the triangularity
of the matrix $\hat{\gamma}\!:\ =\{\gamma_{nk}\}$.
The eigenvalues $\gamma_n=\gamma_{nn}$ are identical with the
flavor nonsinglet anomalous dimensions known from deep inelastic scattering
(moments of the Gribov-Lipatov-Altarelli-Parisi kernel).
In leading order the solution of  (\ref{opcorgeq}) is given by
\begin{eqnarray}
\label{opco-lo}
\left\langle 0| O_{n}(Q^2)|P\right\rangle\ =\
\left({{\alpha_s (Q_0^2)} \over{ \alpha_s (Q^2)}}\right)^{\!\! \gamma_n^{(0)}/
\beta_0}
\ \left\langle0| O_{n}(Q_0^2)|P\right\rangle\ ,
\end{eqnarray}
$$\phantom{MMMMM}
\alpha_s (Q^2)\ =\  { {4\pi} \over
 {\beta_0 \ln\left( {Q^2 / \Lambda^2}\right) } }\ ,
\vspace{6pt}
$$
\noindent
where $Q_0$ is an appropriate reference momentum, $\Lambda$ is the QCD scale
parameter,
\begin{eqnarray}
\label{adcoev1l}
\gamma_n^{(0)}\ =\
C_F\left[  \,3\
+{ 2 \over {(n+1)\,(n+2)}}\
 -\ 4\ \sum^{n+1}_{i=1}\!^{^{\scriptstyle \prime }}\   {1\over i}\ \right]\ ,
\end{eqnarray}
 $C_F=4/3$ and $\beta_0 = (11/3)\, C_A-(2/3)\, n_f$,  with $n_f$ is the
number of active quarks and $C_A=3$.

Since $\gamma_n^{(0)}<0$ for $n>0$,
$\left\langle 0|O_{n}(Q^2)|P\right\rangle^{\rm red}$ decrease
[see Eq.\ (\ref{opco-lo})] with increasing $Q^2,$ so that all harmonics
with $n>0$ will also be suppressed.
Furthermore,  current conservation implies $\gamma_0^{(0)}=0$
so that from Eq. (5)  the asymptotic distribution amplitude follows:
\begin{eqnarray}
\label{daas}
\varphi^{as}(x)\ =
 \lim_{Q^2\to\infty}\ \varphi(x,Q^2)\ =\ 6(1-x)\ x\ ,
\end{eqnarray}
which does not evolve in leading order.
\eject

In next-to-leading order the operators mix under renormalization with each
other. Thus the evolution of
$\left\langle 0|O_{n}(Q^2)|P\right\rangle^{\rm red}$
is determined by an infinite coupled first-order differential equation system.
 Since the anomalous dimension matrix is triangular,
this system can be perturbatively solved,  resulting
in a behavior qualitatively different than the solution from leading order.

For instance, if  the initial condition is set as
$\varphi(x,Q^2_0)=\varphi^{as}(x)=6(1-x)x$
 at the reference momentum square
$Q^2_0$, then  all higher harmonics will  also be excited.
In the limit $Q^2\to \infty$,  these excitations  disappear,
 returning to $\varphi^{as}(x)$.
This effect is investigated more generally and quantitatively in the
following two sections.


\section{Next to Leading Analysis for Fixed Coupling Constant}

To see the essential features of the next-to-leading-order correction,
consider first the solution of the evolution equation for fixed
coupling constant $\alpha_s$. In this case,  the mentioned excitation of
higher harmonics by evolution will not disappear in the asymptotic
limit \hbox{$Q^2 \to \infty$.}
Expansion of $\varphi(x,Q^2)$ with respect to the eigenfunctions
$\varphi_{n}^{ef}(x,\alpha_s)$ of the evolution kernel $V(x,y,\alpha_s)$
provides immediately the solution of the evolution equation:
\begin{eqnarray}
\label{evoeqfi}
\varphi\left(x,Q^2\right)\ =\ \sum_{n=0}^{\infty}\!^{^{\scriptstyle \prime }}\
                                 \varphi_{n}^{ef}(x,\alpha_s)
           \left({{Q^2} \over {Q^2_0}}\right)^{\!\!\gamma_n(\alpha_s)/2}
            \left\langle 0|O_{n}(Q_0^2)|P\right\rangle^{\rm  red}\ .
\end{eqnarray}

The next-to-leading-order corrections to the evolution enters as a two--loop
contribution of the eigenvalues $\gamma_n(\alpha_s)/2$ and as $\alpha_s$
corrections to the eigenfunctions $\varphi_{n}^{ef}(x,\alpha_s)$.
The two--loop corrections of the eigenvalues are well known
from the next-to-leading-order
analysis of deep inelastic scattering \cite{FloRosSac77}.
A closed expression for the $\alpha_s$ corrections to the eigenfunctions
can be derived from conformal constraints and a one--loop calculation of
the special conformal anomaly in Ref.~\cite{DM94} (Here, the result
 is re-expressed by a linear combination of
Lerch transcendent $\phi[x,1,i]$,  and taking into
account the term proportional to~$\beta_0$.),
\begin{eqnarray}
\label{ef}
\varphi_n^{ef}(x,\alpha_s)\ =\
       (-1)^n\  {{2\left(3+2n\right)}\over\left(n+1\right)!}\
       {d^n\over dx^n}\  x^{1 + n}\  {\left( 1 - x \right) }^{1 + n}\
    \left( 1+{\alpha_s \over 2\pi}\  F_n(x)+O(\alpha_s^2)\right)\ ,
\end{eqnarray}
\eject

\noindent
where
\begin{eqnarray}
\label{efcoef}
\!\!\!F_n(x)\ &=&\  {-(\gamma^{(0)}_n-\beta_0)} \left[ {1\over 2}\
{\ln\Big( x(1-x)\Big)}
                               -{\psi}(2+n)+{\psi}(4+2n)\right]
\nonumber\\[.2in]
&&+\ C_F \left[\ {\ln^2\left( {{\textstyle 1-x}\over \textstyle  x}\right)
\over 2}
\ -\   \sum_{i = 1}^{1 + n}
     \left( -{1\over i} + {1+\delta_{0n}\over {2 + n}} \right) \,
      \Big( {\phi}(1 - x,1,i) +
        {\phi}(x,1,i)  \Big)\right.
\nonumber\\[.2in]
&&\left.\qquad\qquad\   +\ 2\left(
    {{\left( 3 + 2\,n \right) \,
     \Big( {\gamma_E} + {\psi}(2+n) \Big) }\over
{\left( 1 + n \right) \,\left( 2+n\right) }}+{\psi}'(2 + n) -
{\pi ^2\over 4} \right)\right]\ ,
\end{eqnarray}
where $\psi(z)=d\ln(\Gamma[z])/dz$, $\gamma_E=0.5772,\,\ldots\, ,$ and
$\phi(x,1,i)=\sum_{k=0}^\infty x^k/(i+k)$.
The term proportional to $\gamma_n^{(0)}$ in Eq.\ (\ref{efcoef}) can be
obtained directly by assuming a nontrivial fixpoint $\alpha_s^\ast$, i.e.,
$\beta(\alpha_s^\ast)=0$, from a conformal operator product expansion
\cite{BroDarFriLep86}.  I thus  refer to it as conformal symmetry predicted
part. Conformal symmetry breaking by the $\beta$-function provides a shift of
the anomalous dimensions $\gamma_n^{(0)} \to \gamma_n^{(0)}-\beta_0$.
The remaining term in Eq.~(\ref{efcoef}) is proportional to the color
factor $C_F$,  and can be interpreted as an `additional' conformal symmetry
breaking term that comes from the renormalization of the conformal operators
in gauge field theory.

\subsection{Corrections to the eigenfunctions}

Consider the asymptotic limit $Q^2\to\infty$.
 As in leading order, the asymptotic distribution amplitude is completely
determined by the eigenfunction $\varphi_0^{ef}$
\begin{eqnarray}
&&\varphi_0^{as}(x,\alpha_s)=\varphi_0^{ef}(x,\alpha_s)\ ,
\nonumber\\[.135in]
\label{phi0ef}
                        &&{\quad\  =}\ 6(1-x)x
\Bigg(1+{\alpha_s\over 4\pi}
\left\{ C_F\left[\ln^2\left({{1-x}\over x}\right)
                               +2 -{\pi^2\over 3}\ \right]
+ \beta_0 \left[\ln\Big( (1-x)x\Big)
                               +{5\over 3}\ \right]\right\}\Bigg)\ .
\end{eqnarray}
The term  in $\varphi_0^{ef}$  proportional to $\beta_0$,
gives a logarithmic modification.
It is very interesting that the conformal symmetry breaking
term provides an
unexpected $\ln^2$ modification of the endpoint behavior.
The $\alpha_s$ correction to the asymptotic distribution amplitude
(\ref{phi0ef}) is shown in Figs.~2(a,b).

We next study quantitatively the $\alpha_s$ contributions for the
eigenfunctions with arbi\-trary~$n$.
For this purpose,  it is technically more convenient to deal
 with the following representation~\cite{DM94}:
\begin{eqnarray}
\label{def-efsum}
\varphi_{n}^{ef}(x,\alpha_s)&=& {(1-x)x\over N_n} \
C_n^{3/2}(2x-1)
+ {\alpha_s \over 2\pi}\  \varphi_{n}^{ef(1)}(x) + O(\alpha_s^2)\ ,
\nonumber\\[.135in]
\varphi_{n}^{ef(1)}(x) &=&
\sum_{k=n+2}^{\infty}\!^{^{\!\!\!\!\!\!{\scriptstyle \prime}}}\quad
 {(1-x)x\over N_k}\
 C_k^{3\over 2}(2x-1)\ c_{kn}^{(1)}\ ,
\end{eqnarray}
where
\begin{eqnarray}
\label{def-efsumco}
c_{kn}^{(1)}&=&
{{(2n+3)\,(\gamma_{n}^{(0)}-\beta_0+4 A_{kn})}\over {(k-n)\,(k+n+3)}} +
{{2(2n+3)\  \Big( A_{kn}-\psi(k+2)+\psi(1)\Big)}\over{ (n+1)\ (n+2)}}\ ,
\nonumber\\[.08in]
\noalign{and}
\nonumber\\[-.125in]
A_{kn}&=&C_F\left[\psi\left({k+n+4\over 2}\right)
-\psi\left({k-n\over 2}\right)
       +2\psi\left(k-n\right)-\psi\left(k+2\right)-\psi(1)\right]
\end{eqnarray}
are only nonzero if $k-n$ even.
To comprehend these $\alpha_s$ contributions quantitatively,
consider the amplitude at $x=0.5$.
However, since the $\ln^2\Big( (1-x)/x\Big)$ term in Eq.~(13)
disappears at $x=0.5$, it is clear that
the large contributions of the endpoint region will be dropped out.
 Nevertheless, from Eq.\ (\ref{def-efsum}) and
$C_{2n}^{3/2}(0)= (-1)^{(n)}\,\Gamma(3/2+n)\,/\,
 \Big(\Gamma (1+n)\,\Gamma (3/2)\Big)$ \cite{BatErd53defC},
the relative contributions
$r_n^{(1)}=\varphi_n^{(1)ef}(0.5)\,/\,
\varphi_n^{(0)ef}(0.5)$
increase logarithmically with $n$, and  are of order
2 for $n=10$ $(\beta_0=0)$, respectively, for $n=2$
$(\beta_0=9)$.

To take into account the missed logarithmic modification in the endpoint
behavior,  it is more reasonable to use the following
quantitative measure for the $O(\alpha_s)$ contribution:
\begin{eqnarray}
\label{sophCon}
R_n(\alpha_s)\ =\
\left(\int_0^1 {N_n\varphi^{ef}_n(x,\alpha_s)^2
\over{\left(1-x\right)\,x}} -1\right)^{\!\!1/2}
    =  {\alpha_s\over 2\pi}\ R_n^{(1)} + O(\alpha_s^2)\ .
\end{eqnarray}
Figure~1(a,b) shows that this analysis provides qualitatively the
same $n$-dependence as for $r_n^{(1)}$, and that the $\alpha_s$
contributions are now larger.
Moreover, the following are common features of the $\alpha_s$
corrections to the eigenfunctions:

\begin{itemize}
\item
For $n=0$ and $\beta_0=0$, only the `additional'
 conformal symmetry breaking
part gives a contribution, of order $\alpha_s/2\pi$.
For $\beta_0\not=0$, this term is partly cancelled.

\item
Contributions from the symmetry predicted and breaking parts have
different phases,  so that the net-contribution is smaller.

\item
In the case of $\beta_0=0$, the  minimum is at n=6.
For $\beta_0\not=0,$ this effect is washed out.
For small $n$ and $\beta_0=0$,  the corrections are small.

\item
The relative corrections are  growing logarithmically,\\
$\ r_n^{(1)}\ \sim\  0.347\beta_0-(2.71+1.39\ln(2+n))\,C_F$\ , \\[.125in]
$R_n^{(1)}\ \sim\ \Big[0.411\beta_0^2+
 \Big( 54.7-35.9\,\ln(2+n)+6.58\,\ln^2 (2+n)\Big)\,C_F^2$ \\
$ {\qquad+}\ \Big(-8.98+3.29 \ln(2+n)\Big)\beta_0
                                           \,C_F\Big]^{(1/2)}$\ , \\
and in the limit $n\to\infty$,
the relative corrections are independent of $\beta_0$.
\end{itemize}

Later, the evolution of $\varphi(x,Q^2)$ will be computed numerically.
For this purpose, it is necessary to know how well the partial sums
\begin{eqnarray}
\label{def-efappr}
\varphi_{ni}^{ef(1)}(x)\ =\
\sum_{k=n+2}^{n+2i}\!^{\!\!^{\!\!{\scriptstyle \prime} }}\
            {(1-x)x\over N_k}\  C_k^{3/ 2}(2x-1)\ c_{kn}^{(1)}
\end{eqnarray}
approximate the functional series (\ref{def-efsum}). This is
also important for the case of running coupling, where the partial waves beyond
the leading order are given by the functional series that have convergence
properties similar to the series for the eigenfunctions.
The relative deviation from $\varphi_{n}^{ef(1)}(x)$ can be measured by
\begin{eqnarray}
\label{def-reldev}
\Delta_{ni}=\sqrt{1-{\textstyle\left.\int_0^1\ dx\
\varphi_{ni}^{ef(1)}(x)^2\right\slash
\Big(x(1-x)\Big)\over
   \textstyle    \left.  \int_0^1\ dx\ \varphi_{n}^{ef(1)}(x)^2\right\slash
\Big( x(1-x)\Big)}}\ ,
\qquad
              \int_0^1dx\  {\varphi_{ni}^{ef(1)}(x)^2\over x(1-x)}\ =
               \sum_{k=n+2}^{n+2i}\!^{\!\!^{\!\!{\scriptstyle \prime} }}\
{\Big( c_{kn}^{(1)}\Big)^2\over N_k}\ .
\end{eqnarray}

Numerical computation shows that for $n=0$, where $\beta_0=0$, the deviation
is $43\%$ for $i=1$, about $10\%$ for $i=5,$ and about $1\%$ for $i=21$.
In general, to get the same deviation for $n>0$,  a larger number
of terms is
taken into account;  e.g., for $n=4$ the deviation is
$50\%$ for $i=5$, $10\%$ for $i=19$, and  $1\%$ for $i=82$.
In Figs.~1(c,d),  the $n$-dependence of the deviation
 is shown for the cases that keep  (c) two terms  and (d) ten terms  of
 the expansion (\ref{def-efappr}).
To remain under the $3\%$ level for $n\le500$ it is necessary to keep $50$
terms. The asymptotic expansion of $\Delta_{ni}$ for large $i$ and $n$, where
$i\ll n$,
\begin{eqnarray}
\label{def-Delta}
\Delta_{ni}\ \simeq\   \sqrt{ {1\over {1+i}} }\
  \sqrt{{{{
  { \Bigg(0.5 -{{\textstyle 2C_F\left[ 2.96+\ln(1+i)\right]{\vphantom{)_)^)}}}
    \over
    {\textstyle \beta_0 -  C_F\left[0.692-4\ln(2+n)\right]}}}}
                      \  \Bigg)^{\!\!\scriptstyle 2}{\vphantom{_)}}}}
\over
    {\;  0.411 +{{\textstyle 48.7 C_F^{\vphantom{^)_)}2}
                                    -8.42C_F\Big[\beta_0
-C_F [ 0.692-4\ln(2+n)]\ \Big]_{\vphantom{\int}}
    \over
    {{{\textstyle \Big(\beta_0-C_F
           \left[ 0.692-4\ln(2+n)\right]\ \Big)}^{\vphantom{)^)}2}}}}}} }
\  +\, \cdots
\end{eqnarray}
is proportional to $1/\sqrt{1+i}$ for fixed $n$.
Furthermore, $\Delta_{ni}$ increases with $n$ and has the limit
$\lim_{n\to\infty} \Delta_{ni}\simeq 0.78/\sqrt{1+i}$.
In this limit there are much larger values when $n$ is moderately large;
e.g., $n\sim 100$:
\begin{eqnarray}
\label{}
\lim_{n\to\infty} \Delta_{n2}\simeq 0.49\ ,\quad
\lim_{n\to\infty} \Delta_{n10}\simeq 0.235\ ,\quad
\lim_{n\to\infty} \Delta_{n50}\simeq 0.11\ ,\quad
\lim_{n\to\infty} \Delta_{n5000}\simeq 0.011\ .
\end{eqnarray}

To approximate the logarithmic endpoint behavior of
$\varphi^{ef}_n(x,\alpha_s)$,
 a much larger number of terms than suggested from
the previous analysis should be taken into account.
For situations where the endpoint behavior is crucial, e.g.,
 for the next-to-leading-order
analysis of the elastic pion form factor, it is better to use the
following integral representation \cite{DM94}:
\begin{eqnarray}
\label{def-efintre}
\varphi_n^{ef}(x,\alpha_s)\ =\
\int_0^1 dy\
 \Big(\ \delta(x-y)+{\alpha_s\over 2\pi}\ c^{(1)}(x,y)+ \,\cdots\,\Big)\
 {(1-y)y\over N_n}\  C_n^{3/ 2}(2y-1)\ ,
\end{eqnarray}
where
\begin{eqnarray}
\label{def-efintc}
c^{(1)}(x,y) & = & (I-{\cal P})
\left({\beta_0\over 2}\ S(x,y)
         -\int_0^1dz\ S(x,z)\  V^{(0)}(z,y)+[g(x,y)]_+\right)\ ,
\nonumber\\[.125in]
{[g(x,y)]}_+ & = & g(x,y) - \delta(x-y) \int_0^1 dz\ g(z,y)\ ,
\nonumber\\[.125in]
g(x,y) & = & C_F \,\theta(y-x)\  {\ln\left(1-{x\over y}\right)
\over (x-y)}\ +\ \left\{ {x \to 1-x \atop y \to 1-y} \right\}\ .
\end{eqnarray}
Furthermore, the convolution with the kernel $S(x,y)$ generates a shift of
the Gegenbauer polynomial order
\begin{eqnarray}
\label{dewf-shiftop}
\int_0^1 dy S(x,y)\ {(1-y)y\over N_n}\
C_n^{3/2}(2y-1)\ =\ {d\over d\rho}  {\Big( (1-x)x\Big)^{1+\rho}\over N_n}\
C_n^{3/2+\rho}(2x-1)_{\,\left\vert_{\scriptstyle \rho=0}\right.} \ ,
\end{eqnarray}
and the operator ${\cal P}$ projects on the diagonal part of the expansion of a
function $f(x,y)$ with respect to $C_i^{3/2}$; i.e.,
${\cal P} f(x,y)=\sum_{i=0}^{\infty} (1-x)x/N_i C_i^{3/2}(2x-1) f_{ii}
C_i^{3/2}(2y-1)$,
where $f_{ij}$ with $0\le i,j\le\infty$ are the expansion coefficients.
Although the operator $\cal{P}$ and the kernel $S(x,y)$ are only defined
implicitly, Eq.\ (\ref{def-efintc}) is nevertheless helpful to convolute
$c^{(1)}(x,y)$ with a given hard scattering amplitude.

Finally, from Eq.\ (\ref{def-efintre}),  the $\alpha_s$ corrections
to  the eigenfunctions can be written as convolution
\begin{eqnarray}
\label{corecef-conv}
\delta^{ef}\ \varphi(x,Q^2)\ =\ {\alpha_s\over 2\pi\ }\int_0^1
    dy\ \left( c^{(1)}(x,y)+ \,\cdots\,\right)\   \varphi^{d}(y,Q^2)\ ,
\end{eqnarray}
where the partial waves of $\varphi^{d}(x,Q^2)$ are given as Gegenbauer
polynomials
\begin{eqnarray}
\label{def-dadiag}
\varphi^{d}(x,Q^2) = \sum_{n=0}^{\infty}\!^{^{\!\scriptstyle \prime }}\
          {(1-x)x\over N_n}\  C^{3/2}_n(2x-1)
       \left({{Q^2} \over {Q^2_0}}\right)^{\!\!\gamma_n(\alpha_s)/2}
                          \left\langle0|O_{n}(Q_0^2)|P\right\rangle^{\rm red}.
\end{eqnarray}
A further advantage of the representation (\ref{corecef-conv}) is that the
above mentioned excitation of higher harmonics is now completely included in
the kernel $c^{(1)}(x,y)$.

\subsection{Corrections to the eigenvalues}

The two-loop corrections to the anomalous dimensions $\gamma_n(\alpha_s)$
are given in Ref.\ \cite{FloRosSac77}.
As in one-loop order $\gamma^{(1)}_n < 0$ for all $n>0$ holds true.
Thus, if these two-loop corrections are resumed in
$$(Q/Q_0)^{\ \textstyle (\alpha_s/2\pi )\,\gamma^{(0)}_n
                                +(\alpha_s/2\pi )^2\,\gamma^{(1)}_n}\ ,$$
the (modified) partial waves for $n>0$ will be more strongly suppressed
than in leading order.
The relative two-loop corrections to $\gamma_n(\alpha_s)$ are about
$4.5 \alpha_s/(2\pi)$ [$4 \alpha_s/(2\pi)$] for all $n>0$ and $n_f=3$
[$n_f=4$], giving a correction of 20\%  for reliable values of
 $\alpha_s\sim 0.35.$ The relative correction
to the evolution of the distribution amplitude is probably of the same order.
(This kind of correction does not appear directly in the evolution of the
asymptotic distribution amplitude, so that in this case they are much
smaller.)

If the corrections arising from the eigenvalues are expanded with respect
to $\alpha_s$, it is possible to write these corrections as convolution with
the leading order solution of the evolution equation,
\begin{eqnarray}
\label{corecev-conv}
\delta^{ev}\ \varphi(x,Q^2)=\left({\alpha_s\over 2\pi}\right)^2
 \ln\left({Q^2\over Q_0^2}\right)
     \int_0^1 dy\  V^{d(1)}(x,y)\ \varphi^{LO}(y,Q^2)\,
\end{eqnarray}
where $V^{d(1)}(x,y)={\cal P}V^{(1)}(x,y)$ is the diagonal part of
$V^{(1)}(x,y)$.

Although the kernel $V^{(1)}(x,y)$ is known in a closed form, it seems a more
difficult task to extract the diagonal part $V^{d(1)}(x,y)$.
A reasonable approximation can be found from the fact that $\gamma^{(1)}_n$
grows like $\gamma^{(0)}_n$, i.e., only logarithmically, for increasing $n$.
The simple form of the asymptotic expansion
\begin{eqnarray}
\label{gammaas0}
\gamma^{as(0)}_n &\simeq& -5.3333\ln (2 + n)+ 0.9215\ ,
\nonumber\\[.125in]
\gamma^{as(1)}_n &\simeq& -(33.237-2.963 n_f)\ln(2+n) + 15.315 - 1.4363 n_f\ ,
\end{eqnarray}
and the eigenvalue equation [which is known from the one-loop approximation of
$V(x,y)$]
\begin{eqnarray}
\label{}
\int_0^1dy\ \big[ v_b(x,y) \big]_{+}\ (1-y)\,y\  C_n^{3/2}(2y-1) & =
&  2\Big(1-\gamma_E-\psi(n+2)\Big)\,  (1-x)\,x\ C_n^{3/2}(2x-1)\ ,
\nonumber\\[.125in]
\big[ v_b(x,y)\big]_{+} & = &  v_b(x,y) - \delta(x-y) \int_0^1 dz\ v_b(z,y)\ ,
\\[.125in]
v_b(x,y) & = &  \theta (y-x)\  {x\over {y(y-x)}}
 +\
\left\{ {x \to 1-x \atop y \to 1-y} \right\}\ ,\nonumber
\end{eqnarray}

\noindent
where $\psi(n+2) = \ln(n+2) + O(1/n)$ for large $n$,
allows us to reexpress Eq.\ (\ref{corecev-conv}) as
\begin{eqnarray}
\label{corecev-conv1}
 \delta^{ev}\,  \varphi(x,Q^2) = \!\bigg({\alpha_s\over 2\pi}\bigg)^{\!\! 2}\,
\! \ln\!\Bigg( {Q^2\over Q_0^2}\Bigg)
\bigg[\int_0^1 dy\,  \Big(a \delta(x-y)+b [v_b(x,y){]}_{+}\Big)\,
\varphi^{LO}(y,Q^2)+R(x,Q^2)\,\bigg]\ ,
\end{eqnarray}
where $a\simeq 0.6315- 0.0918 n_f$
and   $b\simeq 8.309-0.7408 n_f$.
The terms in the sum representation of the remainder
\begin{eqnarray}  \label{}
\!\!R(x,Q^2) &=& \int_0^1 dy \left(V^{d(1)}(x,y) -a \delta(x-y)-
                              b [v_b(x,y){]}_{+}\right)\  \varphi^{LO}(y,Q^2)\
,
\nonumber\\[-.125in]
 \\[.1in]
 &=&{1\over 2}\ \sum_{n=0}^{\infty}\!^{^{\scriptstyle \prime }}\
 {(1-x)x\over N_n}\  C_n^{3/2}(2x-1)
         \left(\gamma^{(1)}_n -\gamma^{as(1)}_n\right)
        \Bigg({{Q^2} \over {Q^2_0}}\Bigg)^{\!\!\alpha_s\gamma_n^{(0)}/(4\pi)}
        \left\langle 0|O_{n}(Q_0^2)|P\right\rangle^{\rm red}\nonumber
\end{eqnarray}
are additionally suppressed by $O(1/n)$.
Thus, for the same accuracy,  the approximation of $R(x,Q^2)$ by a partial sum
requires less terms than the approximation of $\varphi^{LO}(y,Q^2)$ itself.

\subsection{Complete next-to-leading order corrections}

In the asymptotic limit, each given distribution
amplitude $\varphi(x,Q_0^2)$ at reference momentum square
 $Q^2_0$ extends into the asymptotic distribution amplitude (\ref{phi0ef}).
 Thus, in this limit, the relative next-to-leading-order  correction
$\left[\varphi^{NLO}(x)-\varphi^{LO}(x)\right]/\varphi^{LO}(x)$
is uniquely given by
\begin{eqnarray}
\label{relascor}
{\varphi^{asNLO}(x)-\varphi^{asLO}(x)\over\varphi^{asLO}(x)}=
{\alpha_s\over 4\pi}\,
 \Bigg(C_F\bigg[\ln^2\bigg( {1-x\over x}\bigg)+2-{\pi^2\over 3}\,\bigg] +
           {\beta_0} \bigg[ \ln \Big( (1-x)x\Big)+{5\over3}\,\bigg] \Bigg)\ ,
\end{eqnarray}
 so it is large and enhanced in the endpoint region.
The next-to-leading-order contribution by the evolution of the
distribution amplitude is also important  away from this asymptotic
limit.

It is  possible to  get information about the distribution amplitude
at low momentum transfer;  e.g., $Q_0\sim 0.5$ GeV, from nonperturbative
methods such as sum rules \cite{SumRule} and lattice calculation \cite{LatCal}.
However, the obtained results are inconclusive, so it is not possible to
distinguish between the following parameterizations:
\begin{eqnarray}
\label{phiin}
\varphi^{as}(x)\ &=&\ 6x(1-x)\ ,\nonumber \\[.1in]
\varphi^{CZ}(x)\ &= &\ 30x(1-x)\ \Big( 1-4x(1-x)\Big)\ , \\[.1in]
\varphi^{co}(x)\ &=&\ {8\over \pi}\  \Big( x(1-x)\Big)^{1/ 2}\ . \nonumber
\end{eqnarray}
The function $\varphi^{co}(x)$
which was used for the next-to-leading-order analyses of the pion form factor
in Ref.\ \cite{DitRad81} is only one example of further convex amplitudes.
(For a numerical calculation,
 $\varphi^{co}(x)$ is more suitable than broader amplitudes,
 which had previously been assumed to be more realistic.)
Furthermore,  it is assumed that the evolution of $\varphi(x,Q^2)$ for
$Q > 0.5$ GeV can be obtained from the perturbative solution of the
evolution equation.

The evolution of $\varphi(x,Q^2)$ is controlled by Eq.\ (\ref{evoeqfi}), where
the reduced expectation values
$\left\langle 0|O_{n}(Q_0^2)|P\right\rangle^{\rm red}$ are computed
from the nonperturbative input $\varphi(x,Q_0^2)$, which is assumed to be
one of the functions in Eq.\ (\ref{phiin}).
 It follows from Eqs.\ (\ref{evoeqfi}) and (\ref{def-efsum})
up to corrections of order $O(\alpha_s^2)$,
\begin{eqnarray}
\label{inconfi}
\left\langle 0|O_{n}(Q_0^2)|P\right\rangle^{\rm red} &=&
              m_n(Q_0^2) - {\alpha_s \over 2\pi}\
 \sum_{i=0}^{n-2}\!^{^{\scriptstyle \prime }}\
              c^{(1)}_{ni}\  m_i(Q_0^2)\ ,
\nonumber\\
m_n(Q_0^2)&=&\int_0^1 dx\  C_n^{3/2}(2x-1)\  \varphi(x,Q_0^2)\ ,
\end{eqnarray}
where the coefficients $c^{(1)}_{ni}$ are defined in Eq.\ (\ref{def-efsumco}).

Taking into account a sufficient number of terms in the series (\ref{evoeqfi}),
the distribution amplitude  at the factorization scale for exclusive processes
 assumed to be $Q\sim 2$ GeV can be obtained numerically.
The number of active flavors is three, and the value for the fixed coupling
constant is $\alpha_s=0.5$.
The distribution amplitude was approximated by the first 100 nontrivial
terms $(i=0,2,\,\ldots\, ,200)$. The corresponding eigenfunctions
$\varphi^{ef}_n(x,\alpha_s)$ take into account the $(102-i/2)$ terms of the
expansions with respect to Gegenbauer polynomials.
The distribution amplitude
   $\varphi(x_j,Q^2)$ at $Q=2$ GeV was then computed for different points
$x_{j}$, $j=0,1,\,\ldots\, ,70$ and interpolated to a smooth function.

It can be seen in Figs.~2(b,c,d)  that the relative next-to-leading-order
 corrections have the following features:
\begin{itemize}

\item
Independent of the shape of $\varphi(x,Q_0)$,  the relative
next-to-leading-order corrections are
characterized by logarithmic enhancement at the endpoints caused by
both corrections to the eigenfunctions and to the eigenvalues.

\item
For partial waves with $n>0$,  the corrections coming from the eigenvalues are
larger than from the eigenfunctions.
However,  these corrections disappear in the asymptotic limit.

\item
Amplitudes enhanced  at the endpoints, also have larger
relative next-to-leading-order corrections that are negative.
\end{itemize}

Although it was possible for the chosen distribution amplitudes to compute
the evolution in next-to-leading-order numerically, this will be a difficult
task for amplitudes that are broader.
In addition, the complete next-to-leading-order analyses for an exclusive
 process can be done more conveniently if the next-to-leading-order
correction is written as a convolution,  with the distribution amplitude
 $\varphi^{d}(x,Q^2)$ defined in Eq.\ (\ref{def-dadiag}),
which also evolves smoothly in next-to-leading-order
 (no excitation of higher harmonics).
Since $\varphi(x,Q^2)= \varphi^{d}(x,Q^2) +\delta^{ef} \varphi(x,Q^2)$
 from Eq.\ (\ref{corecef-conv}),
\begin{eqnarray}
\label{evol-conv1}
\varphi(x,Q^2)&=& \int_{0}^{1} dy\ \bigg(\delta(x-y) + {\alpha_s \over 2\pi }\
c^{(1)}(x,y) + \cdots\bigg)\  \varphi^{d}(y,Q^2)\ .
\end{eqnarray}
Again, the excitation of the higher partial waves is completely included in the
convolution with $c^{(1)}(x,y)$.
 Notice that  $\varphi^{d}(y,Q_0^2)$ may be used
instead of  $\varphi(y,Q_0^2)$  as an initial condition.
In fact, this corresponds to the choice of another factorization scheme for the
considered exclusive process (redefinition of the soft and hard parts).

The complete $\alpha_s$ correction to the evolution of the distribution
amplitude in next-to-leading-order can easily be obtained from
  (\ref{corecef-conv}) and  (\ref{corecev-conv}):
\begin{eqnarray}
\label{corecal-conv}
\!\!\varphi(x,Q^2)&=&\varphi^{LO}(x,Q^2) + \delta^{ef}\, \varphi(x,Q^2) +
                                        \delta^{ev}\, \varphi(x,Q^2)\ ,
\\[.125in]
&=& \int_{0}^{1} dy \Bigg(\delta(x-y) + {\alpha_s \over 2\pi }
\, \bigg[\,c^{(1)}(x,y) + {\alpha_s \over 2\pi }
\ln\bigg({Q^2\over Q^2_0}\bigg)\   V^{d(1)}(x,y)\,\bigg]
+\cdots\,\Bigg) \varphi^{LO}(y,Q^2)\, .
\nonumber
\end{eqnarray}

\section{Next-to-Leading-Order Analysis for Running Coupling Constant}

This section discusses the solution of the evolution
equation in next-to-leading-order
 for running  coupling, which was derived in \cite{DM94,BroDarFriLep86},
\begin{eqnarray}
\label{daso}
\varphi\left(x,Q^2\right)\ =\
                           \sum_{n=0}^{\infty}\!^{^{\scriptstyle \prime }}\
                                 \varphi_{n}\Big(x,\alpha_s(Q^2)\Big)\,
     \exp \bigg[{1\over 2}\int_{Q^2_0}^{Q^2}\ {dt\over t}
\  \gamma_n\Big( g(t)\Big)\bigg]
                \   \left\langle 0|O_{n}(Q_0^2)|P\right\rangle^{\rm red}\ .
\end{eqnarray}
The partial waves $\varphi_{n}(x,\alpha_s(Q^2))$ are now $Q^2$ dependent
nonpolynomial functions,  known as  functional series
\begin{eqnarray}
\label{daso2l}
\varphi_{n}\Big( x,\alpha_s(Q^2)\Big)\ &=&
{(1-x)x\over N_n}\  C_n^{3\over 2}(2x-1)
             + {\alpha_s(Q^2) \over 2\pi}\  \varphi_n^{(1)}(x,Q^2)+\cdots\ ,
\nonumber\\[.125in]
\varphi^{(1)}_{n}(x,Q^2)\ &=&
  \sum_{k=n+2 }^{\infty}\!\!\!\!^{^{\scriptstyle \prime }}\ \
                          {(1-x)x\over N_{k}}\ C_{k}^{3\over 2}(2x-1)\
      s_{kn}\Big(\alpha_s(Q^2)\Big)\ c^{(1)}_{kn}\ ,
\end{eqnarray}
where $c^{(1)}_{kn}$ are the expansion coefficients of the eigenfunction
defined in Eq.\ (\ref{def-efsumco}) and
\begin{eqnarray}
\label{def-s}
s_{kn}\Big(\alpha_s(Q^2)\Big)\ =\
     {{\gamma_k^{(0)}-\gamma_{n}^{(0)}}\over
               {\gamma_k^{(0)}-\gamma_{n}^{(0)}+\beta_0}}\
\Bigg[ 1-\left( {\alpha_s(Q_0^2) \over\alpha_s(Q^2)}
\right)^{_{\!\!\textstyle 1+(
\gamma_k^{(0)}-\gamma_{n}^{(0)})/ \beta_0}}\Bigg] \ .
\end{eqnarray}
Since these partial waves satisfy the convenient initial condition
\begin{eqnarray}
\label{parwaveincon}
\varphi_{n}\Big(x,\alpha_s ( Q_0^2)\Big)\ =\
{(1-x)x\over N_{k}}\ C_{k}^{3/ 2}(2x-1)\ ,
\end{eqnarray}
the expectation values $ \left\langle0|O_{n}(Q_0^2)|P\right\rangle^{\rm red}$
can be now simpler computed as for fixed $\alpha_s$
\begin{eqnarray}
\label{}
 \left\langle 0|O_{n}(Q_0^2)|P\right\rangle^{\rm red}\ =\
    \int_0^1 dx\   C_{n}^{3\over 2}(2x-1)\ \varphi(x,Q_0^2)\ .
\end{eqnarray}

Because of the asymptotic behavior of $\gamma_k^{(0)}=-4\ln(k+2)+\cdots$,
$s_{kn}\Big( \alpha_s(Q^2)\Big)$ approaches $1$ for $k\gg n$ and
$\alpha_s(Q_0^2)>\alpha_s(Q^2)$ [see Fig.~3(a)].
Consequently, the behavior of
 $\varphi_{n}^{(1)}\,\Big(x,\alpha_s(Q^2)\Big)$ in
the endpoint region is determined by $c^{(1)}_{kn}$; i.e., it has the same
logarithmic modification as $\varphi_{n}^{ef(1)}(x)$.

To avoid this excitation of higher harmonics (Gegenbauer polynomials)
 by the evolution, a new distribution amplitude
 analogous to the case for the  fixed coupling constant
is introduced  that satisfies a diagonal evolution equation
(the corresponding evolution kernel has to be diagonal with respect to
Gegenbauer polynomials).
A formal representation for this transformation kernel was given in
\cite{MiRa86},
\begin{eqnarray}
\label{rot-kern}
W=\int_0^\infty dt \exp{\left\{-(\beta_0-V^{(0)})t\right\}}
\otimes \left[({\cal I-P})V^{(1)}\right]
                                     \otimes\exp{\left\{-V^{(0)}t\right\}}\ ,
\end{eqnarray}
but, as was  pointed out,   this representation cannot be used for
explicit calculations.
Hopefully, changing the factorization scheme for the  exclusive
process under consideration will allow us to factorize
  the process amplitude in terms of the
desired diagonal distribution amplitude  $\varphi^{d}(x,Q^2)$.

For the numerical study of the next-to-leading-order
 corrections, assume that the distribution amplitude at
$Q_0=0.5$ GeV ($\Lambda^{(3)}=0.4$ in next-to-leading-order;
i.e., that $\alpha_s(Q_0^2)\sim 0.9$) can be parametrized
by one of the functions in Eq.\ (\ref{phiin}).
The amplitudes are evolved to a scale $Q=2$ GeV, where $\alpha_s(Q^2)\sim 0.3$.
The number of active flavors is three,  taking into account the first
100 nontrivial terms in the partial sums for both series
(\ref{daso}) and (\ref{daso2l}) [the asymptotic
(Chernyak-Zhitnitsky) distribution amplitude requires only 1~(2)  term(s) in
(\ref{daso})].
The result in Figs.~3(b,c,d)  shows that the relative
 next-to-leading-order corrections for running
coupling have  qualitative and quantitative features similar to those
 in the case
of fixed coupling discussed in Section~3.3.

\section{Summary and Conclusion}

This paper has shown that the (relative)  next-to-leading-order
 correction to the evolution of the pion distribution amplitude
 is rather large, especially in the endpoint region,
and that in this region the negative corrections are larger for enhanced
amplitudes.
The $\alpha_s$ correction to the partial waves comes from the off-diagonal
matrix elements of $\gamma_{nk}$; it can be interpreted as
excitation of higher harmonics (Gegenbauer polynomials) by evolution,
and appears as $\ln(x(1-x))$ and $\ln^2(x/(1-x))$ terms.
The two-loop contribution to the anomalous dimension $\gamma_n$ is
for $n>0$ much larger than the off-diagonal matrix elements
 of $\gamma_{nk}$;  i.e.,  about 20\% of the one-loop approximation.
However, the exponentiation of the two-loop contribution
provides a larger suppression of the corresponding harmonic's as in leading
order (expansion with respect to $\alpha_s$ provides a large (negative)
excitation of the harmonics).

The obtained large next-to-leading-order correction seems to contradict
 a previous analysis  \cite{MiRa86}, where it was found that this correction
is rather small.  The explanation for this discrepancy is that
(1)  only the first few expansion
coefficients $c_{nk}$ were taken into account,
and (2) the authors looked only to the evolution of $\varphi(x,Q^2)$
at $x=0.5$.
Furthermore, the reference momentum chosen  for use in Ref.~[24]  was
$Q_0=10 \Lambda^{(3)}=1$ GeV
in leading order; i.e., $\alpha_s(Q_0^2=1\mbox{GeV}^2)\sim 0.3$.
Such a choice provides a much smaller next-to-leading-order
correction for $Q_0=1.25 \Lambda^{(3)}$ (because
$\alpha_s(Q_0^2=1.25^2 \Lambda^2)\sim 0.9$, perturbation theory should
be valid for the evolution of the distribution amplitude).
 Using a popular parameterization at lower reference
momentum (e.g., $Q_0\sim0.5$ GeV) provides logarithmic correction, which should
be included in the input amplitude at a higher reference momentum.

The  question of whether to include an $\alpha_s$ suppressed logarithmic
correction to the input amplitude $\varphi(x,Q_0^2)$ can be avoid by
chosing a distribution amplitude that evolves smoothly,
 with no excitation of  higher harmonics by evolution.
The amplitude $\varphi^d(x,Q^2)$ satisfies an evolution equation where the
corresponding evolution kernel $V^d(x,y)$ is diagonal with respect to
Gegenbauer polynomials.
Consequently, in such a factorization scheme, the contribution
responsible for the mentioned excitation of higher harmonics is now
included as the $\alpha_s$ correction to the hard scattering amplitude
of the considered process.

Because of the size of the discovered correction
 and its dependence upon the input amplitude,
the evolution of the distribution amplitude has to be included in
the next-to-leading-order analysis of exclusive hard momentum processes.
For large enough $Q^2,$ the Sudakov suppression can be neglected
so that, using the known expressions for the hard scattering amplitudes
of the pion transition form factor and the  electromagnetic form factor,
 it should be straightforward to re-analyze the
next-to-leading-order corrections for these  processes.
Because of the large number of Feynman diagrams,
 the $\alpha_s$ correction to the hard scattering amplitude for
the  $\gamma\gamma\to M^+ M^-$ processes for the case of equal
momentum sharing was only computed  numerically.
  It should nevertheless be possible to estimate the size of the correction
coming from the evolution of the distribution amplitude.
A general next-to-leading-order analysis for arbitrary distribution amplitudes
requires an analytical calculation of the hard scattering amplitude
(448 diagrams).

\section*{Acknowledgments}

For helpful discussions I am grateful to S.~Brodsky, A.~Brandenburg, T.~Hyer,
and O.~Jacob. This work was supported in part by Department of Energy contract
DE--AC03--76SF00515 (SLAC), and by Deutschen Akademischen Austauschdienst.


\newpage

\section*{Captions}
\vspace{.5in}
\begin{enumerate}
\item [] {\bf Fig.  1.}~  Values of $r_n^{(1)}$ (boxes) and $R_n^{(1)}$
(circles) are given when (a) $\beta_0$ is set to zero,
and (b) $\beta_0=9$.
The difference of the relative value of the conformal symmetry
 predicted part (upper half plane)
and of the relative value of the `additional' conformal
symmetry breaking term (lower half plane) is shown by filled boxes (circles).
The lines represent the corresponding asymptotic expressions.
(Subasymptotic terms were also taken into account
 for the approximation of $R_n^{(1)}$.)
The relative deviation of the partial sums $\varphi_{nk}^{ef(1)}(x)$
from the exact $\alpha_s$ correction
$\varphi_n^{ef(1)}(x)$ is given  (c) for  $k=2$ and  (d) for $k=10$.
\vspace*{.2in}
\end{enumerate}

\begin{enumerate}
\item []  {\bf Fig.  2.}~
Evolution of the pion distribution amplitude for fixed
 $\alpha_s=0.5$ and
three active flavors.
As nonperturbative inputs, three distribution amplitudes defined in
Eq.~(32) are chosen at the reference momentum scale
$Q_0=0.5$~GeV.
They are shown in (a); $\varphi^{as}(x,Q^2)$ in leading order (solid),
$\varphi^{CZ}(x,Q^2)$ (dashed), $\varphi^{co}(x,Q^2)$ (dash-dotted),
$\varphi^{as}(x,Q^2)$ in next-to-leading-order (dotted).
The relative  next-to-leading-order corrections at $Q=2$~GeV are shown
 for $\varphi^{as}(x,Q^2)$  in (b),
for $\varphi^{CZ}(x,Q^2)$ in (c), and  for $\varphi^{co}(x,Q^2)$ in (d)
 showing that the endpoint behavior of the distribution amplitudes
changed more drastically under evolution.
The next-to-leading-order corrections of the eigenvalues are
neglected for the dashed line, expanded with respect to $\alpha_s$
for the dash-dotted line, and
taken into account by resummation for the solid line.
The correction in the asymptotic limit is dotted.
\vspace*{.2in}
\end{enumerate}

\begin{enumerate}
\item  [ ]  {\bf Fig.  3.}~  The evolution of the pion distribution amplitude
for running $\alpha_s$ is essentially determined
 by the matrix valued function $s_{kn}\Big( \alpha_s(Q^2)\Big) $.
 (a) shows that $s_{kn}\Big( \alpha_s(Q^2)\Big) $ defined in Eq.~(38) as a
function of $k/(n+2)$ is nearly $n$ independent, and for
$\alpha_s(Q^2)=0.5 \alpha_s(Q_0^2)$ it is almost of order $O(1)$.
The relative  next-to-leading-order corrections for $\varphi^{as}(x,Q^2)$
in (b), for $\varphi^{CZ}(x,Q^2)$ in (c), and for $\varphi^{co}(x,Q^2)$ in (d)
are comparable to the fixed coupling result.
Here,  $\alpha_s(Q_0^2)=0.9$, $\alpha_s(Q^2)=0.3$, and
three active flavors were chosen.
The meaning of the solid, dashed, and dash-dotted lines is the same as for
 Fig.~2(b,c,d).
\vspace*{.2in}
\end{enumerate}

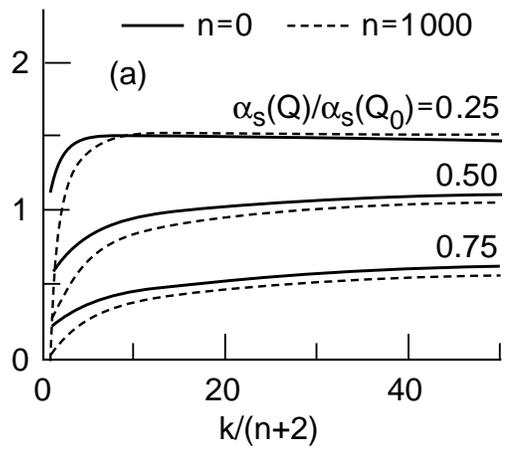
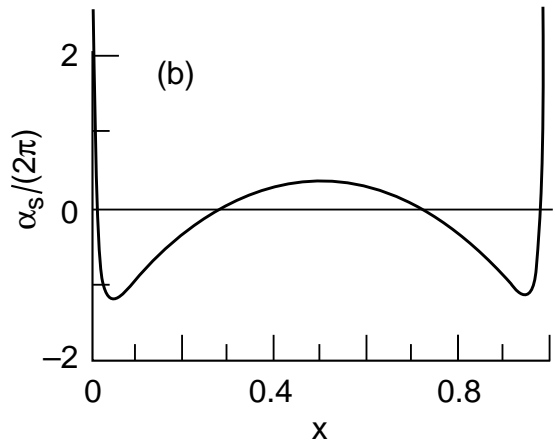
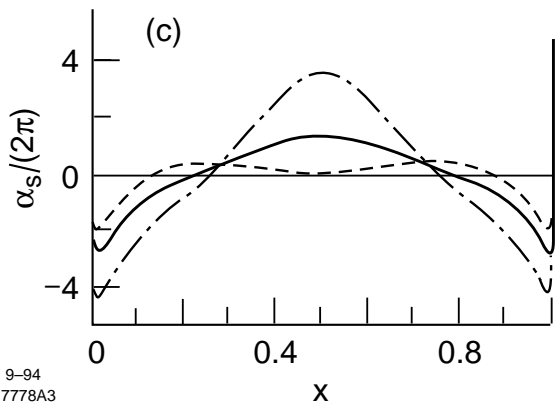
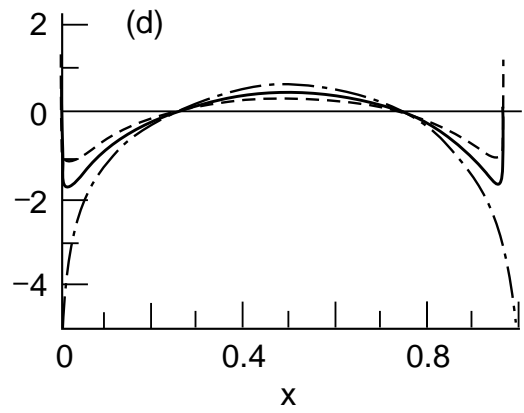
\end{document}